\documentclass[12pt,epsfig]{article}
\usepackage{graphicx,amsmath,amssymb}

\newlength{\dinwidth}
\newlength{\dinmargin}
\def\lapproxeq{\lower .7ex\hbox{$\;\stackrel{\textstyle
<}{\sim}\;$}}
\def\gapproxeq{\lower .7ex\hbox{$\;\stackrel{\textstyle
>}{\sim}\;$}}
\def\gtrsim{\lower .7ex\hbox{$\;\stackrel{\textstyle
>}{\sim}\;$}}
\def\lesim{\lower .7ex\hbox{$\;\stackrel{\textstyle
<}{\sim}\;$}}

\def\be{\begin{equation}}
\def\ee{\end{equation}}
\def\bea{\begin{eqnarray}}
\def\eea{\end{eqnarray}}

\begin{document}                                                    
\titlepage                                                    
\begin{flushright}                                                    
IPPP/09/24   \\
DCPT/09/48 \\                                                    
\today \\                                                    
\end{flushright}                                                    
                                                    
\vspace*{2cm}                                                    
                                                    
\begin{center}                                                    
{\Large \bf Soft physics at the LHC\footnote{Topical Review to be published in J.Phys.G, based on two lectures by Misha Ryskin at the St. Petersburg Winter School, Gatchina, February 2009}}\\   
                                                    
\vspace*{1cm}                                                    
M.G. Ryskin$^{a,b}$, A.D. Martin$^a$, V.A. Khoze$^{a,b}$ and A.G. Shuvaev$^b$ \\                                                    
                                                   
\vspace*{0.5cm}                                                    
$^a$ Institute for Particle Physics Phenomenology, University of Durham, Durham, DH1 3LE \\                                                   
$^b$ Petersburg Nuclear Physics Institute, Gatchina, St.~Petersburg, 188300, Russia            
\end{center}                                                    
                                                    
\vspace*{2cm}

\begin{abstract}
We recall the main features of the Regge approach used to understand soft interactions at LHC and higher energies. Unitarity tames the power growth of the elastic proton-proton scattering amplitude with energy, and leads to the migration of the secondary particles produced in high-energy proton-proton collisions to larger transverse momenta. We discuss, in qualitative terms, the role of processes containing large rapidity gaps (LRG), and the probability that the gaps survive population by secondaries produced in additional soft interactions. We explain how the Regge diagram corresponding to a LRG event simultaneously describes events with different (single, double, etc.) particle density in the same rapidity interval. We show that the role of these, enhanced, multi-Pomeron diagrams can be studied by measuring multiplicity fluctuations and long-range rapidity correlations between secondaries produced at the Tevatron and the LHC. Finally, we make a list of the characteristic features of the multi-Pomeron description of soft interactions that may be observed at the high energies accessible at the Tevatron and the LHC. 
\end{abstract}

\section{Introduction}

``Soft'' physics is manifest in various places in different guises. The interest in soft interactions at {\it very high} energies is stimulated, theoretically, by the fact that the observed hadronic cross sections are growing with energy and, experimentally, by the advent of the LHC. 

At the LHC energy we have to allow for  unitarity effects, which are necessary to provide the consistency of the strong interaction. In fact, when we face a cross section which grows as a power 
of the energy, there are two possible ways to restore unitarity.

First, we may introduce a new particle (sufficiently heavy to have not been observed yet) which cancels the growth of the cross section at energies larger than the mass of the particle (for example, Higgs boson,
super-partners, etc.). After this, we deal with a {\it weak} interaction,
where the main contribution comes from one, or a few, simple diagrams. More complicated Feynman graphs just describe corrections to the lowest-order amplitude.

Another possibility is to consider a {\it strong} interaction. Here, from the beginning, we start with a Hermitian Lagrangian, which should already account for unitarity. When the Born amplitude (or an amplitude obtained via the summation of some group of diagrams) becomes too large and the interaction becomes strong, then new more complicated diagrams must enter the game. It is the contribution due to these new graphs which tames the growth of cross section such that the final result satisfies unitarity.

 In the case of QCD, where the BFKL amplitude ($A\propto s^{1+\Delta}$) grows as a non-integer power ($\Delta \propto \alpha_s$) of energy, $\sqrt{s}$, we expect to observe the second scenario\footnote{It looks impossible to cancel the non-integer power of $s$ by  introducing a new particle.}. Therefore it is important to observe, and to study, the role of more complicated diagrams (arising from multiple interactions) at LHC energies, and to trace how the theory restores the unitarity
of the {\it strong} interaction.
That is, from a theoretical viewpoint, it is of great interest to observe
experimentally how unitarity tames the growth of high-energy hadronic
amplitudes, leading to saturation both in the transverse momenta $k_t$ and in the impact parameter
$b$ distributions of the produced particles. In other words, unitarity replaces the growth
of the amplitude by the growth of configuration space occupied by the particles, both in $b$ and in $k_t$.

Here, we do not present new results but attempt to recall, using modern language, the ideas and understanding of high-energy soft interactions which originated some 40 or more years ago.
The discussion may be divided into three main topics: (i) elastic scattering and the total cross section,
(ii) the cross sections of processes with gaps in rapidity (including high-mass diffractive dissociation), and, (iii) the saturation of particle densities in transverse momenta $k_t$.

Soft high-energy $pp$ interactions are clearly important at the LHC. Moreover, they can complicate our ability to observe new physics. For example, topic (iii) is relevant to the search for new physics,
since particles with a rather large $k_t$ from the underlying inclusive event will affect the jet searching algorithm, as is already the case for jets at HERA and the Tevatron.  On the other hand, the evaluation of the cross section for an exclusive process, such as $pp \to p+{\rm Higgs}+p$, requires knowledge of the gap survival probability $S^2$,
which is the subject of topic (ii), see, for example, \cite{epip,albrow,all,kaid08}.

Here, we focus on the qualitative features of high energy soft interactions. These features are quite general. With the advent of the LHC, it is timely to gather them together and to emphasize the underlying physics.

\section{Total and elastic cross sections  \label{sec:tot}}

The behaviour of scattering amplitudes in the high energy, $\sqrt{s}$, small momentum-transfer squared, $-t$, domain is well described by Regge theory; that is, by the singularities of amplitudes in the complex angular momentum, $j$, plane, see, for example, \cite{regge,bkk}. For instance, the measured $\pi^- p \to \pi^0 n$ amplitude behaves as
\be
A(s,t) \propto s^{\alpha_{\rho}(t)}
\ee
where the $\rho$-trajectory, $j=\alpha_{\rho}(t) \simeq 0.5 +0.9t$ (with $t$ in ${\rm GeV}^2$), passes through the spin-1 $\rho$-meson resonance in the `crossed' $t$-channel $\pi^- \pi^0 \to \bar{p}n$; that is, $\alpha_{\rho}(t=m^2_{\rho})=1$.

\begin{figure} 
\begin{center}
\includegraphics[height=7cm]{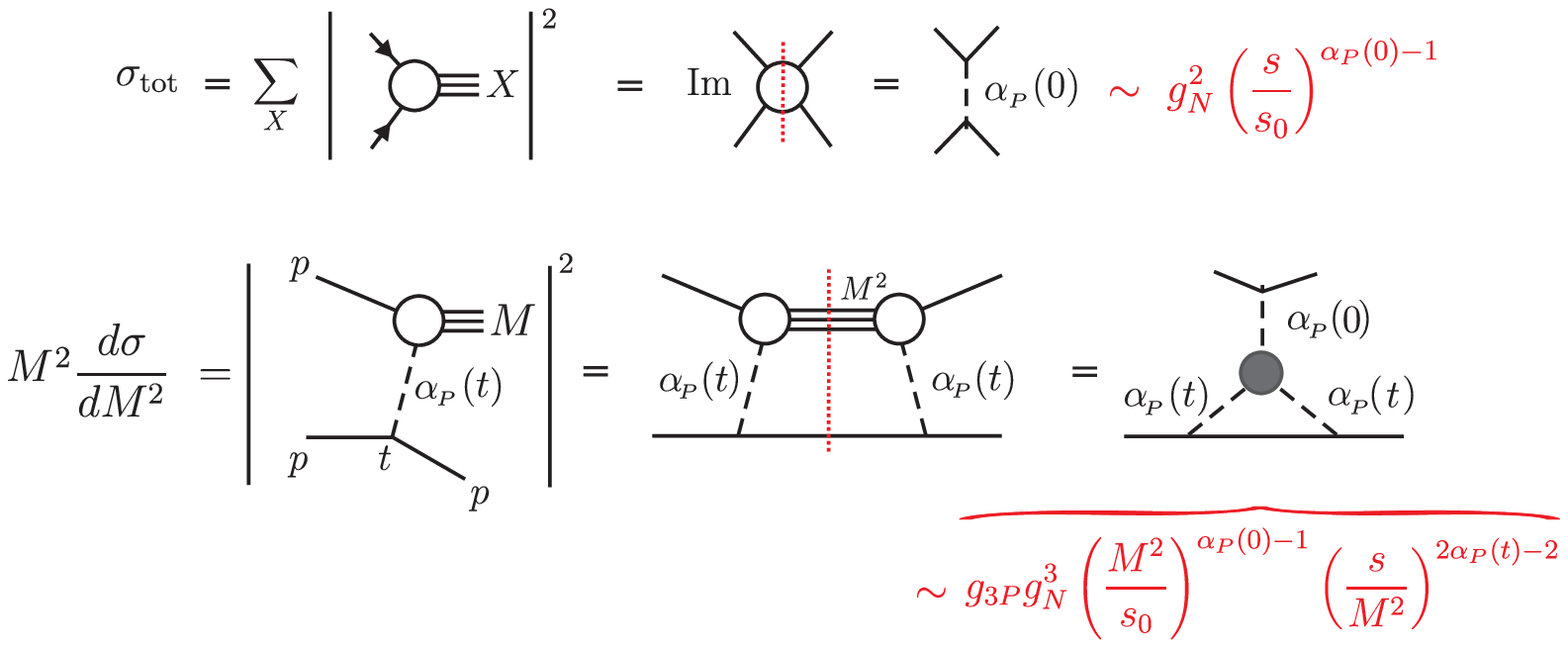}
\caption{(a) A sketch of the optical theorem which, via unitarity, relates the $pp$ total cross section to the imaginary part of the forward elastic $pp$ amplitude; (b) the application of the optical theorem to obtain the cross section for the dissociation of the `beam' proton into a high-mass system, see Sections \ref{sec:SD} and \ref{sec:LRG2}. The high-energy Pomeron-exchange expressions that are shown for the {\it bare} amplitudes have sizeable absorptive (that is, unitarity or screening) corrections; $g_N$ and $g_{3P}$ are the Pomeron-proton and triple-Pomeron couplings respectively. }
\label{fig:OPT}
\end{center}
\end{figure}

On the other hand, total cross sections are observed to grow slowly with energy and are associated with the exchange of a trajectory with vacuum quantum numbers. The simplest possibility is to assume that at high energy these cross sections, such as the $pp$ total cross section  $\sigma_{\rm tot}$, are driven by an isolated pole at $j=\alpha(t)$, which gives an ($pp$) elastic amplitude
\be
A(s,t) \propto s^{\alpha_P (t)},
\ee
and, via the optical theorem of Fig. \ref{fig:OPT}, a total cross section
\be
\sigma_{\rm tot} \propto s^{\alpha_P (0)-1}.
\ee
The pole with the largest intercept, originally with $\alpha_P(0)=1$ since high energy total cross sections were thought to have constant asymptotic behaviour, was called the {\it Pomeron}\footnote{A discussion of the history of the Pomeron is given in \cite{bkk}.}. Here we are interested in Tevatron and higher energies which are sufficiently large to be able to neglect the contributions of all secondary trajectories (which all have intercepts $\alpha (0) \simeq 0.5$).

A popular parametrization of elastic $pp$-scattering amplitude by Donnachie-Landshoff (DL) is the Regge form  \cite{DL}
\begin{equation}
A(s,t)=\eta \sigma_0 F_1^2(t)s^{\alpha_P(t)}
\label{eq:DL}
\end{equation}
where the signature factor $\eta$ gives the complex phase, $F_1$ is the electromagnetic form factor of the proton,  and the {\it effective} soft Pomeron 
trajectory 
\be
\alpha_P=1+\Delta+\alpha' t~\simeq~1+0.08+0.25t,
\label{eq:dl}
\ee 
where $t$ is in GeV$^2$. The intercept $\alpha(0)$ just above 1 reproduces the observed slow growth of the total hadron-hadron cross sections at high energies. Indeed, the effective Pomeron pole amplitude, (\ref{eq:DL}), gives a good description of the total and elastic differential cross section data up to Tevatron energies. However, this simple parametrization is expected to become increasingly deficient at higher energies. Unitarity is an easy way to see this.

To discuss unitarity effects it is convenient to work in impact parameter, $b$, space, since at high energy the position of the particle in $b$ is practically frozen. Thus the value of $b$ determines the orbital momentum $l$ of the incoming proton, 
$l=\sqrt{s}b/2$; that is, fixed $b$ corresponds to a particular partial wave $l$.
In $b$ space the elastic unitarity equation\footnote{The amplitude is normalised to
$\sigma_{\rm tot}=2\int d^2b ~{\rm Im} A(b)$.}
\begin{equation}
2{\rm Im} A(b)=|A(b)|^2+G_{\rm inel}(b)
\label{eq:elun}
\end{equation}
limits the value of ${\rm Im}A \le 2$. However, this limit corresponds to a pure elastic interaction with phase of elastic amplitude, 
\begin{equation}
A_l=i(1-\exp(2i\delta_l)),
\label{eq:el1}
\end{equation}
 equal to $\delta_l=\pi/2$. Normally at high energies
the inelastic contribution $G_{\rm inel}$ dominates, leading to a large inelasticity; that is, to a large imaginary part, ${\rm Im} ~\delta  \gg 1$. In this so-called `black disk' limit\footnote{In general, an amplitude with $1<{\rm ~Im} A<2$ is not forbidden. The corresponding model, so-called $U$-matrix unitarisation, was discussed in \cite{TT}. However, such a model which asymptotically leads to a pure elastic interaction, without any particle production, does not look probable.}, we have ${\rm Im} A\to 1$. Recall that, if we were to neglect the small contribution from the real part, then Im$A(b)$ can be measured directly from experiment as the Fourier transform
\be
{\rm Im}~A(b)~=~\int \sqrt{\frac{d\sigma_{\rm el}}{dt}~\frac{16\pi}{1+\rho^2}}~J_0(qb)~\frac{qdq}{4\pi},
\ee
where $q^2=|t|$, $J_0$ is a Bessel function and $\rho^2 \equiv ({\rm Re}A/{\rm Im}A)^2<0.02$.

\begin{figure} [t]
\begin{center}
\includegraphics[height=10cm]{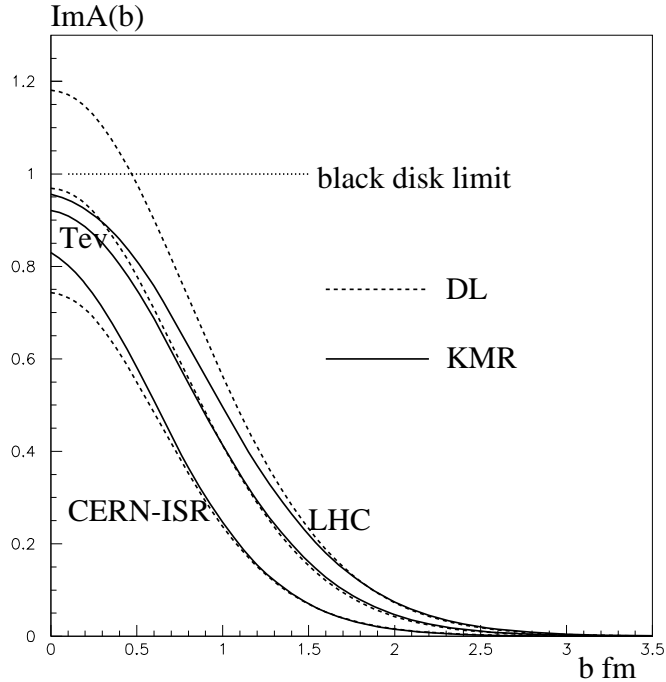}
\caption{The impact parameter profiles of the imaginary part of the elastic amplitude for both the DL effective Pomeron pole \cite{DL} and the KMR multi-Pomeron \cite{KMR} parametrizations, for collider energies $\sqrt{s}=62.5,~1800, ~14000$ GeV relevant to the CERN-ISR, the Tevatron and the LHC respectively.}
\label{fig:dl}
\end{center}
\end{figure}

Note that the DL amplitude $A(b=0)$ crosses the black disk limit between the Tevatron and LHC energies, see Fig.~\ref{fig:dl}. Thus we expect that at the LHC the cross
section will be less than that given by the DL parametrization, while the slope of elastic cross section, $B_{\rm el}$, should be larger. The partons occurring in the proton wave function will be pushed
away from the centre to the periphery. This suppression of the amplitude at low values of $b$ automatically increases the elastic slope $B_{\rm el}$, since $B_{\rm el} \propto R^2 = \langle b^2 \rangle$ where $R$ is the interaction radius.   

For this reason the value of $\alpha'=0.25$ GeV$^{-2}$ in (\ref{eq:dl}) should not be considered as the slope of the bare Pomeron trajectory. Part of this slope is generated by the stronger absorptive corrections at smaller $b$. Indeed, the shrinkage of the diffractive cone observed in deep-inelastic scattering at HERA (say, in $\rho$-meson or in $J/\psi$ diffractive production) is smaller than that observed in $pp$-scattering.

Actually to satisfy elastic unitarity it is sufficient to consider an 
eikonal model which sums up the exchanges of any number of Pomerons between the incoming protons. The model gives a result analogous to (\ref{eq:el1}) 
\begin{equation}
A(b)=i(1-\exp(-\Omega(b)/2))
\label{eq:eik}
\end{equation}
where, up to a normalization factor, the opacity $\Omega(b)$ plays the role of the scattering phase $\delta_l$. In the eikonal model, (\ref{eq:eik}), it is the opacity $\Omega\propto s^{\alpha_P-1}$ which is now described (or parametrised) by Pomeron exchange, rather than the final amplitude $A$. The situation is sketched symbolically in Fig. \ref{fig:epip}(a).
However this is not the whole story.
The eikonal model does not include events with Large Rapidity Gaps
(LRG), arising from the dissociation of one or both of the protons into high-mass systems, which are observed experimentally. We speak of single or double high-mass diffractive dissociation.
% The description of these events will play a significant role in %our discussion. 
We introduce high-mass dissociation in the next Section and give detailed discussion of these LRG events in Section \ref{sec:LRG}.

Before we do this, let us recall that the interaction radius (i.e. the elastic slope) expected at the LHC will already exceed 1 fm -- the distance where {\it confinement} may enter the game. Thus it is possible that confinement will stop the growth of the radius; that is, the growth of $B_{\rm el}$.
Another interesting possibility, proposed by V.V.Anisovich \cite{Anis},  is that the partons (the gluons and quarks)
in the proton will start to form a few colourless clusters (like the nucleons in nucleus), rather than the normal homogeneous distribution. This may lead to a rich diffractive dip structure in the differential elastic cross section $d\sigma_{\rm el}/dt$. It is an additional argument to study experimentally the $t$ behaviour of the elastic cross section at the LHC\footnote{Reviews of the predictions for the total cross section, and the elastic scattering cross section over an extended $|t|$ interval, can be found in \cite{revt}.}.

\section{Diffractive dissociation \label{sec:SD}}
\begin{figure}
\begin{center}
\includegraphics[height=8cm]{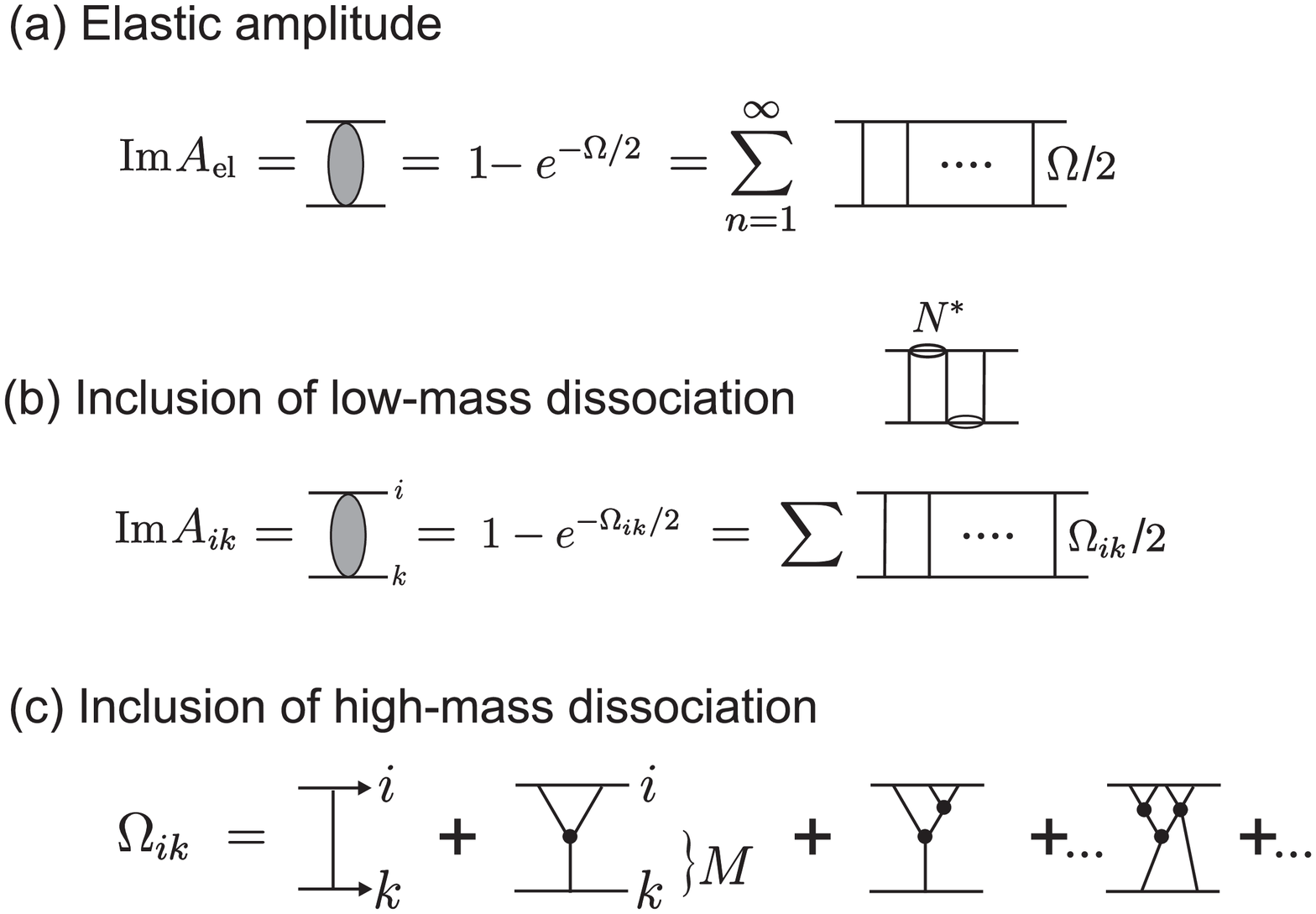}
\caption{(a) The single-channel eikonal description of elastic scattering; (b) the multichannel eikonal formula which allows for low-mass proton dissociations in terms of diffractive eigenstates $|\phi_i\rangle,~|\phi_k\rangle$; and (c) the inclusion of the multi-Pomeron-Pomeron diagrams which allow for high-mass dissociation.  }
\label{fig:epip}
\end{center}
\end{figure}

So much for elastic scattering, which we may call elastic diffraction. Now we turn to inelastic diffraction, which is a consequence of the {\it internal structure} of the protons. Besides the pure elastic two-particle intermediate states shown in Fig.~\ref{fig:epip}(a), there is the possibility of proton excitation, $p \to N^*$, shown in the small sketch in Fig.~\ref{fig:epip}(b). At high energies, when the lifetime of the fluctuations of the fast proton is large, $\tau \sim E/m^2$, the corresponding Fock states can be considered to be `frozen'. Each constituent of the proton can undergo scattering and thus destroy the coherence of the fluctuations. As a consequence, the outgoing superposition of states will be different from the incident particle, so we will have {\it inelastic}, as well as elastic, diffraction.

To discuss inelastic diffraction, it is convenient to follow Good and Walker \cite{GW}, and to introduce states $\phi_k$ which diagonalize the $A$ matrix. These, so-called diffractive, eigenstates only undergo `elastic' scattering. To account for the internal structure of the proton we, therefore, have to enlarge the set of intermediate states, from just the single elastic channel, and to introduce a multi-channel eikonal. The situation is pictured in Fig.~\ref{fig:epip}(b).

What about proton dissociation into high-mass systems? At first sight, it appears that we may account for it by simply enlarging the number of diffractive eigenstates $\phi_k$. Even if this were practical, we would still face the problem of double counting when partons originating from the dissociation of the beam and the `target' protons overlap in rapidities.  Instead, high-mass dissociation is described in terms of so-called ``enhanced'' multi-Pomeron diagrams. The first, and simplest, is the triple-Pomeron diagram, shown in Fig.~\ref{fig:OPT} and again in Fig.~\ref{fig:epip}(c). In fact, high-mass dissociation is much larger than low-mass dissociation at very high energy, and it will play a central role in our discussion of soft interactions.  For simplicity, therefore, we do not discuss low-mass dissociation further, but consider that it can be easily allowed for in terms of a multi-channel eikonal. So, from now on we should regard $\Omega$ as an effective opacity embodying low-mass dissociation.

Why does high-mass dissociation become so important at high collider energies? 
A simplified way to see this is to note that the cross section for dissociation of a proton into a high-mass ($M$) system has the approximate form\footnote{Here, for simplicity, we assume an essentially flat energy dependence, $\sigma \sim s^{\epsilon}$ with $\epsilon {\rm ln}s <1$. The final equality in (\ref{eq:simple}) can be deduced by taking the ratio of the couplings indicated in the Regge expressions in Fig.~\ref{fig:OPT}.}
\be
\sigma_{\rm SD}~=~\int \frac{M^2 d\sigma_{\rm SD}}{dM^2}\frac{dM^2}{M^2}~\sim~\lambda {\rm ln}s ~ \sigma_{\rm el},
\label{eq:simple}
\ee
with $\lambda \equiv g_{3P}/g_N$, where $g_{3P}$ is the triple-Pomeron coupling and $g_N$ is the coupling of the Pomeron to the proton, see Fig.~\ref{fig:OPT}. The ln$s$ `rapidity' factor comes from the integration $\int dM^2/M^2$. Note that here we have used $\sigma_{\rm el} \sim g_N^4$ and $M^2 d\sigma_{\rm SD}/dM^2 \sim g_N^3 g_{3P} \sim \lambda g_N^4$, see Fig~\ref{fig:OPT}. Eq. (\ref{eq:simple}) is grossly oversimplified. For one thing, the dissociation cross section is subject to even greater absorptive (that is, multi-Pomeron) corrections than the elastic cross section. When these are taken into account it is found from analyses of the available high-energy data that
\be
\lambda \equiv g_{3P}/g_N~\simeq~0.25,
\ee
see, for example, \cite{KMR}. The important point is that the relative size of the $\sigma_{\rm el}$ and $\sigma_{\rm SD}$ cross sections is governed, not simply by the parameter $\lambda$, but rather by $\lambda {\rm ln}s \sim 1$. For each fixed rapidity interval the probability of high-mass dissociation is relatively small, but the cumulative effect in the complete interaction amplitude is enhanced by the large phase available in rapidity at high collider energies.

\section{The structure of the Pomeron  \label{sec:Pom}}

In the Regge approach, we have seen that high-energy soft interactions are driven by Pomeron exchange, together with its absorptive (multi-Pomeron) corrections. We may call this the ``soft'' Pomeron. Note that the $s$-channel asymptotic behaviour arising from $t$-channel Pomeron exchange corresponds to an interaction radius which grows with energy, $R^2 \sim {\rm ln}s$. The Pomeron must also describe multiparticle interactions. The sum of ladder diagrams of the type of Fig.~\ref{fig:2}(a) is the simplest multiparticle structure which reproduces the power-like $s^{\alpha}$ behaviour of the Pomeron pole. Before the advent of QCD, ladder diagrams composed of $t$-channel mesons (dominantly spinless pion exchange\footnote{Pion-exchange was considered to be the dominant mechanism due to the proximity of the pion pole, at $t=m^2_\pi$, to the $s$-channel physical domain ,$t \le 0$, due to the smallness of the pion mass.}) were summed, and it proved difficult to generate an intercept $\alpha (0) \ge 1$.

Shortly after the discovery of QCD it was noticed \cite{LN} that (colourless) two-gluon exchange has the properties of Pomeron exchange: vacuum quantum numbers, even signature and a singularity at $j=1$. Due to the polarisation vectors of the spin-one particle (the gluon), the numerator of the propagator contains the $g_{\mu \nu}$ tensor such that the 4-momenta of the incoming fast protons occur in the form $p_a^\mu g_{\mu\nu} p_b^\nu=(p_a\cdot p_b) \simeq s/2$. Therefore the gluon-exchange amplitude contains an extra power of $s$ as compared with spinless particle exchange. 

Later, by applying the Regge approach to gluons, rather than to hadrons, it was possible, if the gluon transverse momenta $k_t$ were sufficiently large, to describe high energy (low $x$) interactions in terms of perturbative QCD. Then, on summing the leading contributions, where the smallness of the QCD coupling $\alpha_s$ is compensated by large values of ln$1/x$, we build up the ``hard'' Pomeron. Below, we briefly trace this history and then apply these ideas to gain insight into high-energy soft interactions.

\subsection{Ladder structure of the Pomeron pole before QCD}
\begin{figure}
\begin{center}
\includegraphics[height=3.8cm]{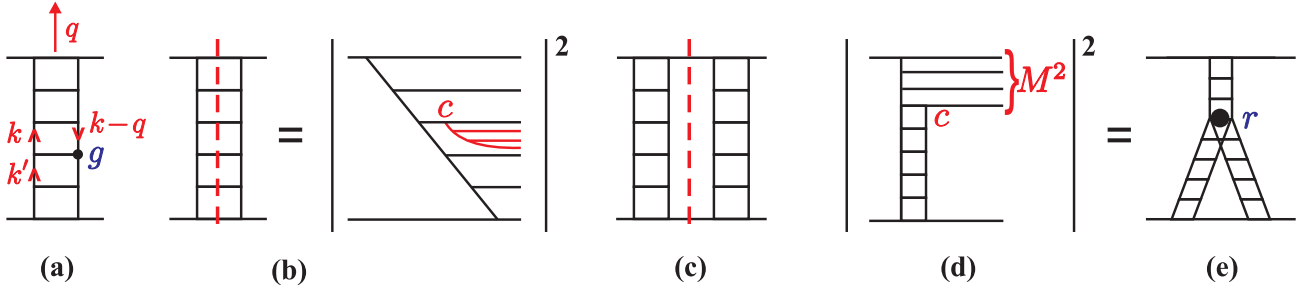}
\caption{Feynman diagrams corresponding to (a) one-Pomeron exchange, (b) the $2 \to n$ cross section, (c) two-Pomeron exchange, and (d,e) the triple-Pomeron contribution to high-mass $M$ diffractive dissociation. The particles that are `cut' by the vertical dashed lines in (b) and (c) are on-mass-shell.}
\label{fig:2}
\end{center}
\end{figure}

In terms of Regge theory, the high-energy amplitude is described by the exchange of a 
Pomeron (or a few Pomerons). As mentioned above, from a quantum field theory viewpoint, the Pomeron may be regarded as the sum of ladder-type diagrams, Fig.~\ref{fig:2}(a). In other words, it corresponds to a sum of completely inelastic
$2\to n$ processes; that is, to the last term $G_{\rm inel}=1-\exp(-\Omega)$ in the unitarity equation (\ref{eq:elun}). Now let us `cut' the Pomeron; see Section \ref{sec:cut} for further discussion. That is, we treat the graph of Fig.~\ref{fig:2}(a), not as a diagram for the amplitude, but rather as the diagram for the $2\to n$ cross section 
\be
\sigma(2\to n)=A^*(2\to n)\cdot A(2\to n),
\ee
 see Fig.~\ref{fig:2}(b). Then, we get the 
inelastic production of $n$ particles homogeneously distributed in the available rapidity, $y$, interval. 

In the original `soft' Regge approach it was assumed that the transverse momenta $k_t$ of these secondaries were limited\footnote{Nowadays, with QCD, as mentioned above and discussed below, the Pomeron is described by a ladder built of the gluons. As a result, due to the dimensionless QCD coupling, there is no parameter to limit the value of $k_t$, and BFKL equation predicts the
`diffusion' of the gluons in $\ln k_t$ space \cite{BFKLdiff}.}.
The original idea, proposed by Amati, Fubini and Stanghellini
in 1962 \cite{AFS}, was to explain a large high-energy cross section, which did not decrease with energy, as a sequence of interactions in which the energy of a pair of adjacent particles (pions) is relatively small and lies in the resonance region where the interaction is strong. The integration corresponding to each cell of such a ladder is over the rapidity $y$ and the transverse momentum $k_t$, see Fig.~\ref{fig:2}(a). If we denote the momentum transferred through the whole ladder by $q_t$, then the contribution from a particular cell is $\alpha(q_t^2)dy$ where
\be
\alpha(q_t^2)~=~ \frac{g^2}{16\pi^2} \int \frac{d^2 k_t}{(k_t^2+m^2)((k_t-q_t)^2+m^2)},
\label{eq:s0}
\ee
and $g$ is the coupling. The integration over the rapidities gives
\be
A(Y)~=~\sum_n \frac{1}{n!}\prod^n_i \int^Y_0 \alpha dy_i~=~\sum_n \frac{(\alpha Y)^n}{n!} ~=~e^{\alpha Y}~=~s^{\alpha},
\ee
where the total rapidity spanned by the ladder $Y={\rm ln}~s$, and $n!$ reflects the identity of the cells or, equivalently, the ordering of the rapidities $y_i$.

A convenient alternative way to obtain this result is to consider the evolution equation in rapidity
\be
\frac{dA(y)}{dy}~=~\alpha ~A(y).
\label{eq:AFS}
\ee
In such a form $\alpha$ plays a role similar to a DGLAP splitting function. 

Note that the value of the `effective spin' $\alpha (q_t^2)$ depends on the momentum transfer through the ladder, $-t=q_t^2$. The slope of the Pomeron trajectory $\alpha '=d\alpha (t)/dt$ at $t=0$ is controlled by the mean value of the transverse momenta $\alpha' \propto g^2/\langle k^2_t \rangle$ or, in the case of (\ref{eq:s0}), by the mass $m$ of the $t$-channel particle.

However, with this procedure it turns out to be practically impossible to obtain a large intercept $\alpha(0)>1$ whilst keeping the $t$-channel exchanges to have masses $m<1-1.5$ GeV. The corresponding couplings $g$ are too small. On the other hand, if we include heavier states, then we go to larger $k_t$; that is to small distances where it looks reasonable to work in terms of perturbative QCD.

\subsection{Ladder structure of the Pomeron pole after QCD \label{sec:PQCD}}

With QCD, the Pomeron, at least at small distances, is described by the ladder of Fig.~\ref{fig:2}(a) built of gluons, rather than of spinless particles; with the coupling $g^2/4\pi$ in (\ref{eq:s0}) replaced by the QCD coupling $\alpha_s$. The integral over $k_t$ now has a more complicated structure. It is convenient to consider the recursion relation between the amplitudes $f_n(x,k_t)$ with $n$ and $n-1$ cells
\be
f_n(x,k_t)~=~\frac{N_c \alpha_s}{\pi}\int^1_x\frac{dx'}{x'}\int \frac{d^2k'_t}{\pi}~K(k_t,k'_t)f_{n-1}(x',k'_t),
\ee
where we keep just the leading ln$1/x'$ term in the $x'$ integration. Note that now we have to account for the dependence of the amplitude on the transverse momenta, $k_t$, of the $t$-channel gluons. The recursion relation may be rewritten in differential form, as the BFKL evolution equation \cite{bfklevol}
\be
\frac{df}{d{\rm ln}1/x}~=~\frac{N_c\alpha_s}{\pi}K\otimes f
\label{eq:bfkleq}
\ee
so that $f \propto x^{-\Delta}$, where
\be
\Delta~=~N_c\alpha_s \langle K \rangle /\pi ~=~ \alpha(t)-1,
\ee
where $\langle K \rangle$ is the leading eigenvalue calculated using equation ($\ref{eq:Kf}$) below. 
Since we now have spin-one gluons in the ladder, the intercept $\alpha(0)=1$ already in the limit $\alpha_s \to 0$. 

In fact, if just the leading ln$1/x$ contributions are summed then, explicitly, $\Delta \equiv \Delta_0 = \bar{\alpha}_s $4ln2, where $\bar{\alpha}_s \equiv 3\alpha_s/\pi$ \cite{bfklevol}. However,  the next-to-leading logarithm (NLL) contributions are numerically rather large. It was found \cite{bfklnlo} that, accounting these $\alpha_s$ corrections, $\Delta \simeq \Delta_0(1-6\bar{\alpha}_s)$. For the relevant, not too small, values of $\alpha_s $, a resummation is necessary. Since the NLL contribution is mainly of kinematic origin, it is possible, and looks reasonable, to resum all the major higher-order corrections.
This leads to $\Delta \simeq 0.3$ \cite{bfklresum} in a wide region of
$k_t$.

For simplicity, we consider just the forward amplitude\footnote{For non-zero $t$ the kernel $K$ is a bit more complicated, but qualitatively similar in form.} with $q_t=0$. Then the elastic forward amplitude is
\be
A=is\int\frac{d^2k_t}{\pi k_t^2}~f(x,k_t),
\ee
and, at LO, the BFKL kernel $K$ acts as
\begin{equation}
K(k_t,k'_t)\otimes f(x,k'_t)~=~\frac 1{(k_t-k'_t)^2}\left\{f(x,k'_t) -
\frac{k^2_t f(x,k_t)}{k'^2_t+(k_t-k'_t)^2} \right\}.
\label{eq:Kf}
\end{equation}
Since in QCD we deal with a dimensionless coupling $\alpha_s$ and a massless gluon, the mean value of $k_t$ in each cell is determined by the value of transverse momentum $k'_t$ in the previous cell. 
This is an important property of the BFKL equation -- that is, the so-called diffusion in $\ln k^2_t$. The value of $L \equiv \ln k^2_t$ in the current cell may differ from that in the previous cell by some
quantity $\delta L\sim 1$. In other words, at each step of
evolution, not only the impact parameter $b$ can be changed by
$\delta b\sim 1/k^2_t$, but also the value of $\ln k^2_t$ can be changed by $\delta L\sim 1$.

\subsection{The transition from the `hard' to the `soft' Pomeron \label{sec:trans}}

 There are phenomenological hints that at large distances the ``soft'' Pomeron should have qualitatively similar structure as the ``hard'' (QCD) Pomeron. Indeed,
 first, no irregularity is observed in the HERA data in the transition region, $Q^2\sim 0.3 - 2 ~{\rm GeV}^2$, between the `soft' and `hard' interaction domains; the data are smooth throughout this region.
Second, a small slope $\alpha'_P \lapproxeq 0.1 ~{\rm GeV}^{-2}$ 
 of the Pomeron trajectory, is obtained in global analyses of all available soft high-energy data, after accounting for absorptive corrections and secondary Reggeon contributions.  This indicates that the typical values of $k_t$ inside the Pomeron amplitude are relatively large ($\alpha'\propto 1/k^2_t$).
 Finally, recent `soft' model data analyses \cite{KMR,GLMM,DN} which account for the enhanced absorptive effects find an intercept of the initial, bare Pomeron $\Delta=\alpha_P(0)-1\simeq 0.3$ close to the intercept of the BFKL Pomeron after the NLL corrections are resummed \cite{bfklresum}.
Thus it looks reasonable to assume that in the soft domain we deal
 with the same perturbative QCD Pomeron;  at least, there is a smooth transition from the soft to the hard Pomeron.

In summary, the {\it bare} hard Pomeron, with a trajectory with intercept $\Delta \equiv \alpha_P(0)-1 \simeq 0.3$ and small slope $\alpha '$, is subject to increasing absorptive effects as we go to smaller $k_t$ which allow it to smoothly match on to the attributes of the {\it soft} Pomeron. In the {\it limited} energy interval up to the Tevatron energy, some of these attributes (specifically those related to the elastic amplitude) can be mimicked or approximated by an {\it effective} Pomeron pole with trajectory intercept $\Delta \equiv \alpha_P(0)-1 \simeq 0.08$ and slope $\alpha '=0.25~{\rm GeV}^{-2}$.

\section{Events with Large Rapidity Gaps  \label{sec:LRG}}

For further discussion, we write the BFKL evolution equation (\ref{eq:bfkleq}) in the oversimplified form
\be
\frac{dA(y)}{dy}~=~\Delta ~A(y).
\label{eq:evoleq}
\ee
where $\Delta=\alpha_P(0) -1$ acts as the splitting function. This evolution equation is the analogue of (\ref{eq:AFS}), but, now, the possibility of multiparticle production leads to an additional power growth of the amplitude, $A(t=0) \propto (s^\Delta ) s$, in comparison with the Born amplitude, $A_B \propto s$. 
In other words, the evolution (\ref{eq:evoleq}) describes the development of the parton cascade from the `beam' proton to the `target' proton, and the $s^\Delta$ growth of the inelastic cross section reflects the power growth
of the parton density in the cascade.

After one of the partons interacts with the target the coherency of the beam
proton wave function is destroyed and a number of secondaries is produced.
Note that in the ladder of Fig.~\ref{fig:2}(b) only one branch of the whole cascade is shown,
which finally leads to a parton being absorbed by the target. The number $N$ of secondaries produced is equal to the number of steps (partons) in this particular branch, shown in Fig.~\ref{fig:2}(b). According to (\ref{eq:evoleq}), where $\Delta$ plays the role of the particle density per unit of rapidity, we have $N\simeq \Delta\cdot Y$. On the other hand, each of the partons continue to develop its own cascade. One example is shown in Fig.~\ref{fig:2}(b), where parton $c$ develops its own branch. However, the branches which were not affected by the target conserve their coherence, and  in the final state appear as a single parton. Thus we get the power growth of the amplitude, but only a logarithmic growth of the multiplicity $N$.  The studies of coherence phenomena in a partonic cascade were pioneered by V.N. Gribov in 1972, see \cite{gribov}.

\subsection{Large Rapidity Gap contribution to the total cross section \label{sec:LRG2}}
The fact that the coherence of the wave function of the beam proton was
destroyed after the interaction with the target, leads not only to inelastic high-multiplicity production, but via unitarity relation  (\ref{eq:elun}), also
to elastic scattering. Elastic scattering is due to the absorption of an initial (coherent) component in some domain of the impact parameter $b$ plane, and originates from the remaining part of the initial wave function which conserves the coherence
between the partons. That is, elastic scattering occurs as the `shadow' of the inelastic interaction. If the inelastic contribution $G_{\rm inel}$ is represented by the single ladder of Fig.~\ref{fig:2}(b), then the elastic contribution may be drawn as a `cut' between the two Pomerons of Fig.~\ref{fig:2}(c), see also Fig.~\ref{fig:4}.

\begin{figure}
\begin{center}
\includegraphics[height=7cm]{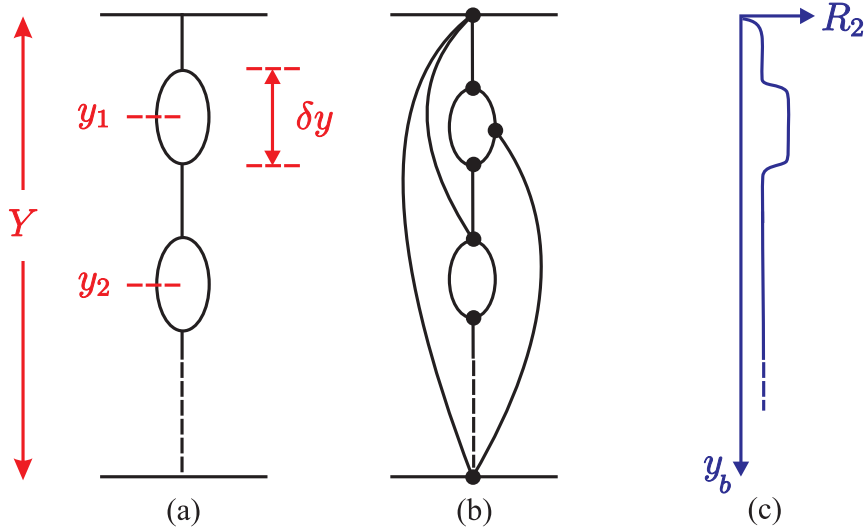} 
\caption{A diagram for a process with several LRG centred at rapidities $y_i$: (a) neglecting the gap survival factor $S^2$, and (b) including the $S^2$ suppression caused by both eikonal (depicted here by the Pomeron line connecting the two protons) and enhanced rescattering involving intermediate partons. Plot (c) shows the expected long-range correlations $R_2(y_a,y_b)$ of (\ref{eq:R2}), corresponding to diagram (a), with $y_a=y_1$, supplemented by eikonal screening.}
\label{fig:3}
\end{center}
\end{figure}

So much for elastic scattering of the incoming proton. However,
an intermediate parton in a ladder may be scattered elastically as well.
A unitarity equation, analogous to (\ref{eq:elun}), may, and should be, written
for any intermediate parton. The elastic scattering of an intermediate parton $c$ can be represented by the diagram of Fig.~\ref{fig:2}(d). Here, both branches of the cascade which start to evolve from parton $c$ save their coherence.
Therefore in the final state we will have a Large Rapidity Gap between the 
parton $c$ and the target. Such a process is called the diffractive dissociation of the beam proton into a high-mass state $M$. It is described by the triple-Pomeron diagram of Fig.~\ref{fig:2}(e), in which one 
Pomeron is split into two Pomerons, see also Fig.~\ref{fig:OPT}. The probability $r(\equiv g_{3P})$ of this splitting within a unit rapidity interval is relatively small. First, due to the small parton $c$
density $\Delta$, where $\Delta$ arises from (\ref{eq:evoleq}). Next, due to the parton-(target) proton cross section being smaller than the proton-proton cross section, since the incoming proton contains many partons. On the other hand, each intermediate parton in the ladder may
generate  a Pomeron splitting. Thus, the whole effect
accumulated during the evolution (\ref{eq:evoleq}) is {\it enhanced} by the parton multiplicity $N$ -- in other words, by the size of available rapidity space $Y$, see also the discussion around equation (\ref{eq:simple}).
Therefore a triple-Pomeron diagram (like that in Fig.~\ref{fig:2}(e)) or diagrams with more complicated multi-Pomeron vertices (like those in Fig.~\ref{fig:3}(b)) are called {\it enhanced} 
diagrams. 

Already at the Tevatron, the cross section of the events with a LRG, that is of single- and double-diffractive dissociation of one or both protons (integrated over the mass $M$ of the final `diffracted states'), is comparable to the elastic $\bar{p}p$ cross
section. If we were to take the contribution of the diagram of Fig.~\ref{fig:2}(e) literally and integrate over the rapidity of vertex $r$, then, at sufficiently high energy, the cross section of events with a LRG would exceed the whole inelastic $G_{\rm inel}$ contribution described by a single 
Pomeron, that is the contribution of  Fig.~\ref{fig:2}(b) (recall $\alpha_P(0)>1$). 

The situation is even worse when we consider events containing a large number of LRG, arising, for example, from the multi-Pomeron diagram of Fig.~\ref{fig:3}(a). For simplicity, let us first account for gaps of limited size, say $\delta y_i<y_0$.  We denote the probability to form such a gap as $\gamma$. Then, integrating over the central position $y_i$ of each gap, we obtain for $Y \gg y_0$ the contribution
$$ \sum_n \frac{(\gamma Y)^n}{n!}\, =\, s^\gamma,$$
which grows as a power of $s$. If we neglect the limit $\delta y_i<y_0$, and allow for the integration over the gap size $\delta y_i$, then  the cross section increases faster than any power of $s$, and clearly violates unitarity. This problem was discussed long ago in \cite{FK}.

Two main scenarios were proposed to prevent such a growth and to restore unitarity. First, the so-called {\it weak} coupling scenario assumes that the
triple-Pomeron vertex $r\propto t$ (and correspondingly the value of $\gamma$) vanishes as $t\to 0$ \cite{GM1}. Since the mean value of momentum transfer $\langle t\rangle$ decreases as $1/\ln s$ (the shrinkage of the diffractive cone) the decrease of $r$ compensates the increase of the available rapidity interval. However this hypothesis is not supported by  experiment\footnote{Still the weak coupling possibility is not rejected completely and it will be important to study high-mass dissociation at the LHC in the low $t$ region to confirm this conclusion; see \cite{LKMR}.}.

In the favoured {\it strong} coupling scenario \cite{GM2}, the multi-gap cross section is
suppressed by a small `gap survival probability' $S^2$ which decreases with energy. In this case the Pomeron coupling $r\to const$ as $t\to 0$,
while the small probability of LRG production is due to the large probability that the gaps are populated by secondaries produced in additional soft rescattering interactions between the protons (and also the intermediate partons).

In the simple eikonal model, (\ref{eq:eik}), the probability, not to have an additional inelastic interaction which will populate the gap, is given by a factor $S^2=\exp(-\Omega)$. Recall that in the eikonal model $G_{\rm inel}=1-\exp(-\Omega)$.  However, this factor accounts for the rescattering of the incoming {\it protons} only. Since, now, we have also rescattering between the {\it intermediate} partons we have to consider
more complicated multi-Pomeron diagrams, like those shown in Fig.3(c).

\subsection{Multi-Pomeron diagrams and the AGK cutting rules \label{sec:cut}}
We emphasize that each multi-Pomeron exchange diagram describes simultaneously a
few different processes. We have already seen that the first ladder diagram gives, on the one hand, the elastic $pp$ scattering amplitude, Fig.~\ref{fig:2}(a), while, on the other hand, it may be considered as the `cut' diagram for the cross section for multiparticle production, Fig.~\ref{fig:2}(b). In the latter case cutting the Pomeron of Fig.~\ref{fig:2}(a), that is taking the discontinuity, disc$A$=2Im$A$, gives the contribution to multiparticle production, $G_{\rm inel}=2{\rm Im}A$. 
 
\begin{figure}
\begin{center}
\includegraphics[height=4cm]{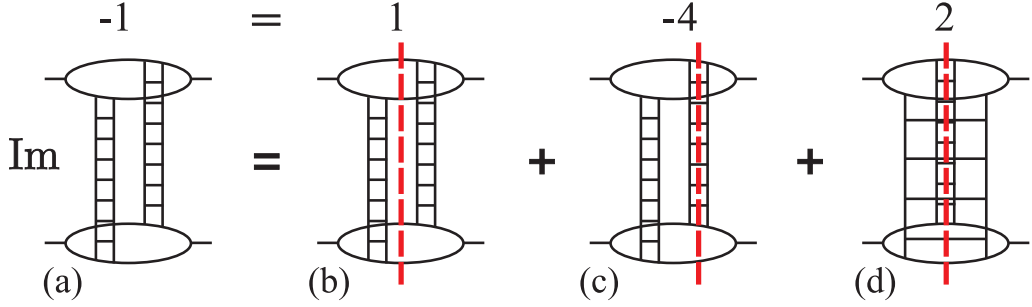}
\caption{Different processes described by cutting the two-Pomeron exchange diagram. The four diagrams correspond, from left to right, to the contribution to the imaginary part of the elastic {\it amplitude}; to the cross section $\sigma_0$ with zero multiplicity in the central region; to the process with single Pomeron multiplicity ($\sigma_1$); and, finally, with double Pomeron multiplicity ($\sigma_2$).}
\label{fig:4}
\end{center}
\end{figure} 
Now consider the two-Pomeron exchange diagram of Fig.~\ref{fig:2}(c). Strictly speaking, the diagram should be drawn more precisely, as has been done
in Fig.~\ref{fig:4}. We have the possibility to cut both Pomerons simultaneously, \ref{fig:4}(d),
to cut only one Pomeron, \ref{fig:4}(c), or to cut between the two Pomerons, \ref{fig:4}(b). The sum of all cuts gives the total contribution of two-Pomeron exchange to the imaginary part of the elastic amplitude. This contribution is negative and describes the {\it absorptive}
correction to the one-Pomeron amplitude. On the other hand, the elastic cross
section corresponding to Fig.~\ref{fig:4}(b) is positive; as is the cross section of the events with a `doubled' particle density, which we obtain by cutting both Pomerons as in Fig.~\ref{fig:4}(d). The only negative contribution is the absorptive correction to the events with a single particle density (i.e. the correction to single-Pomeron exchange). The famous AGK cutting rules \cite{AGK} gives the relation between the different subprocesses originating from the same Reggeon diagram.
For the respective cut diagrams of Fig.~\ref{fig:4}, the relation is 
\be
\sigma_0~:~\sigma_1~:~\sigma_2~=~1~:~-4~:~2,
\label{eq:142}
\ee
 while the whole contribution to the elastic amplitude is $1-4+2=-1$.

Analogous relations hold for more complicated diagrams with many Pomerons. Note that, according to the AGK rules, the multi-Pomeron diagrams do not change the single particle inclusive cross section $d\sigma/dy$. Indeed, the particle density arising from  diagram~\ref{fig:4}(d) is twice larger than that from \ref{fig:4}(c), and these two contributions cancel each other. The same is true for more complicated multi-Pomeron diagrams.

\begin{figure} [t]
\begin{center}
\includegraphics[height=8cm]{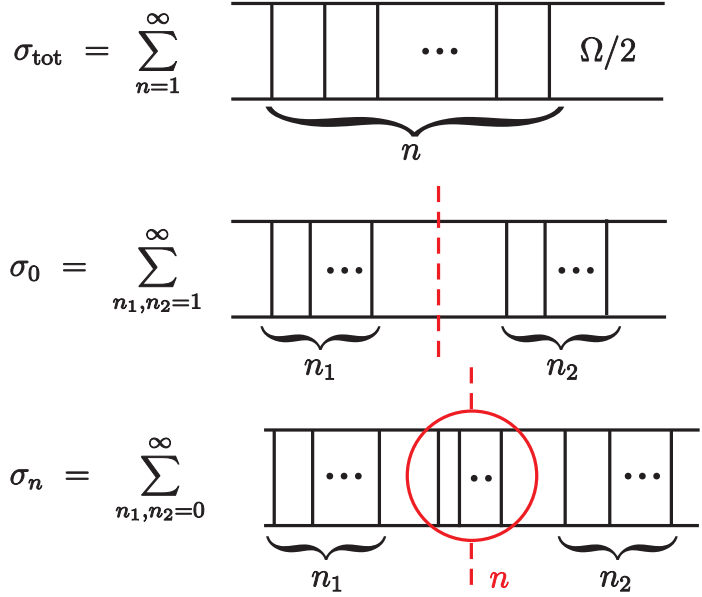}
\caption{The eikonal structure of the total interaction: $\sigma_{\rm tot}(b)=\sigma_{\rm el}(b)+\sigma_{\rm inel}(b)=\sigma_0(b)+\sum_n \sigma_n(b)$. The dashed vertical lines represent the cut leading to the elastic contribution, $\sigma_0$, and the cuts of the $n$ Pomerons in $\sigma_n$. The circle is to indicate that all $n$ Pomerons are cut.}
%there are several different ways to cut the $n$ Pomerons.}
\label{fig:A}
\end{center}
\end{figure}

If we start with the {\it eikonal} model, then the sum of all the multi-Pomeron exchange diagrams gives the total probability of the interaction at fixed $b$,
\be
\sigma_{\rm tot}=2(1-e^{-\Omega/2})=2\sum^\infty_{n=1} (-1)^{n-1}\frac{(\Omega/2)^n}{n!}=\sum^\infty_{n=0} \sigma_n (b),
\label{eq:s1}
\ee
where $\Omega(b)/2$ is given by one-Pomeron exchange and
\be
\sigma_0=\sigma_{\rm el}=(1-e^{-\Omega/2})^2,~~~~~~~~~~~~\sigma_n=\frac{\Omega^n}{n!}e^{-\Omega}~~~~~{\rm for}~~n \ge 1.
\label{eq:pois}
\ee
 Here, $\sigma_0(b)$ is the probability of the elastic 
interaction; and $\sigma_n(b)$ is the probability to produce a particle density $n$ times larger than that arising from one-Pomeron exchange, that is the probability to cut $n$ Pomerons in the whole amplitude. It is easy to see that
\be
\sigma_{\rm inel} \equiv \sum^\infty_{n=1}\sigma_n= \sum^\infty_{n=1}\frac{\Omega^n}{n!}e^{-\Omega}= 1-e^{-\Omega}=G_{\rm inel}
\label{eq:s3}
\ee
is the whole inelastic contribution at a fixed $b$. The penultimate expression for $\sigma_{\rm inel}(b)$ represents the whole probability, 1, minus the probability $e^{-\Omega}$ to have no inelastic interaction, whereas in the previous expression for $\sigma_{\rm inel}$ each term $\Omega^n/n!$ represents the probability of $n$ inelastic interactions (where $n!$ accounts for the identity of the interactions) multiplied by $e^{-\Omega}$ which guarantees that there are no further inelastic interactions. The multi-Pomeron exchange structure of processes of (\ref{eq:s1})-(\ref{eq:s3}) is sketched in Fig. \ref{fig:A}.
 
\begin{figure} [t]
\begin{center}
\includegraphics[height=11cm]{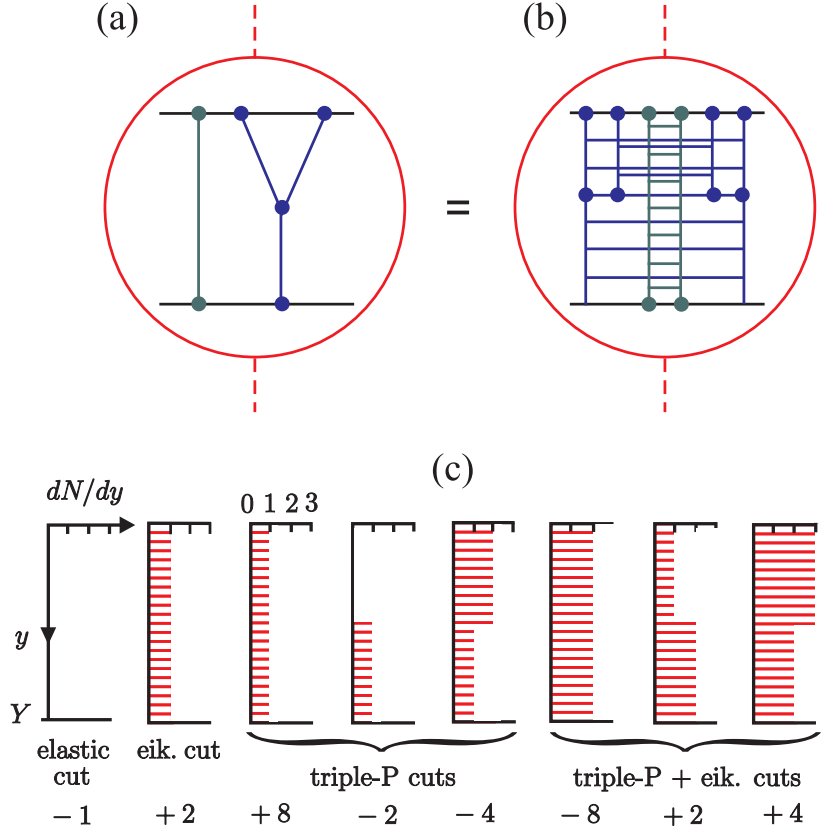}
\caption{(a) A triple-Pomeron diagram screened by one eikonal Pomeron. (b) The gluon ladder structure of the diagram. The dashed line and circle imply that there are several different ways (in fact, 8) to cut this multi-Pomeron diagram. (c) The particle density $dN/dy$, as a function of rapidity $y$, expected for the processes obtained from these 8 different cuts, together with their relative probabilities, which may be obtained from the AGK cutting rules.}
\label{fig:B}
\end{center}
\end{figure}

Similar rules, with the same combinatorics, are valid in the presence of {\it enhanced} screening. At each multi-Pomeron vertex, we can cut one or more Pomerons, or cut between the Pomerons placing the uncut Pomerons either to the left or to the right of the cut. For illustration, we discuss the contributions arising from cutting a triple-Pomeron diagram which is screened by a single eikonal Pomeron, as shown in Fig. \ref{fig:B}(a). The Pomeron ladders corresponding to this diagram, Fig. \ref{fig:B}(b), can be cut in 8 different ways. First, we have the elastic $\sigma_0$ cut between all the ladders. Next, the cut of the eikonal Pomeron ladder gives the usual single-Pomeron multiplicity. Then, we have three possibilities to cut the triple-Pomeron graph; one with single multiplicity in the whole rapidity interval, one with a gap in the upper part, one with double multiplicity in the upper part. Finally, we have the same three, but now simultaneously cutting the eikonal Pomeron. Thus, the different cuts, which lead to 8 different process, give the particle multiplicities as a function of rapidity that are sketched in Fig. \ref{fig:B}(c). The AGK cutting rules may be used to determine the relative probabilities of these processes; the result is given below the multiplicity plots. For the example of Fig. \ref{fig:B} we have to use relation (\ref{eq:142}) twice. For instance, the probability $-8$ comes from $2\times -4$, whereas $-4$ comes from $-2\times 2$ ($-2$ corresponds to the cut of the triple-Pomeron diagram without cutting the single Pomeron; note that in (\ref{eq:142}) $-4=-2-2$, where the two terms correspond to cutting one or the other Pomeron in Fig.\ref{fig:4}). 

Moreover, AGK rules show that, in general, these factors do not depend on exactly how the multi-Pomeron vertices are cut. To be specific, the probability of a particular subprocess is given just by the combinatorial factors to choose $n$ cut Pomerons from $m=n_1+n+n_2$ Pomerons of the original amplitude. A factor 2 comes from disc$A$=2Im$A$ of each cut Pomeron, and another factor 2 comes from the possibility to place any uncut Pomeron to the right or left of the cut.

\subsection{Long-range rapidity correlations}
Note that the eikonal model (\ref{eq:eik}) predicts a
long-range correlation between the secondaries produced in different rapidity intervals. Indeed, we have possibility to cut any number of Pomerons. Cutting $n$ Pomerons we get an event with multiplicity $n$ times larger than that generated by one Pomeron. The probability to observe a particle from a diagram where $n$ Pomerons are cut is $n$ times larger than that from the diagram with only one cut Pomeron. The observation of a particle at rapidity $y_a$, say, has the effect of enlarging the relative contribution of diagrams with a larger number of cut Pomerons. For this reason the probability to observe another particle at quite a different rapidity $y_b$ becomes larger as well. This can be observed experimentally via the ratio of inclusive cross sections
\begin{equation}
R_2~=~\frac{\sigma_{\rm inel}d^2\sigma/dy_ady_b}{(d\sigma/dy_a)( d\sigma/dy_b)}-1~=~
\frac{d^2N/dy_ady_b}{(dN/dy_a)(dN/dy_b)}-1,
\label{eq:R2}
\end{equation}
where $dN/dy=(1/\sigma_{\rm inel})d\sigma/dy$ is the particle density.

Without multi-Pomeron effects the value of $R_2$ exceeds zero only when the
two particles are close to each other, that is when the separation 
$|y_a-y_b|\sim 1$ is not large. Such short-range correlations arise from resonance or jet production. However, multi-Pomeron exchange leads to a
long-range correlation, $R_2>0$, even for a large rapidity difference between the particles, $|y_a-y_b|\sim Y$.
In fact, such a long-range correlation, with $R_2\sim 0.2$ for $|y_a-y_b|\sim 5$, has even already been 
observed in the old CERN-ISR data \cite{R-ISR}.

In the eikonal case, the value of $R_2$ may be calculated explicitly from the Poisson distribution of $\sigma_n$ given in (\ref{eq:pois}). At fixed $b$ it gives
\be
R_2~=~\left(1+\frac{1}{\Omega}\right)(1-e^{-\Omega})-1.
\ee
For a low optical density $(\Omega \ll 1)$ the correlation arises mainly from the two-Pomeron exchange diagram, and increases as $R_2=\Omega/2$. At very large $\Omega$, the Poisson distribution becomes narrow in the number $n$ of cut Pomerons, and $R_2$ decreases as $1/\Omega$. The maximum value is $R_2 \simeq 0.3$ for $\Omega \simeq 1.8$.

It would be informative to study the Pomeron loop structure by observing long-range correlations at the
LHC energy. In the pure {\it eikonal} case, the correlation should be the same throughout the whole available rapidity interval (except, of course, near the edges of rapidity interval, where multiparticle production is affected by energy conservation constraints). When we have  {\it enhanced} diagrams, that is  Pomeron loops which produce LRG which occupy only a part $\delta y$ of the rapidity space, the corresponding correlation takes place only within this $\delta y$ interval. Thus the existence of Pomeron loops may be revealed {\it both} as  LRG events (from the cut with zero multiplicity inside
the $\delta y$ interval), and as long-range $R_2$ correlations in the $\delta y_i$ intervals (from the cuts generating large multiplicities of secondaries). The `range' of the correlation
reflects the size of the Pomeron loop in rapidity space. An example of this behaviour of $R_2$ is shown in Fig.~\ref{fig:3}(c)

\subsection{The decreasing cross section}
Note that the simplest scenario to restore the unitarity at very high energies is not yet completely rejected. We mean the possibility that asymptotically the Pomeron intercept becomes less than 1, 
$\alpha_P(0)<1$, and at very high energies the total cross section starts to decrease with energy.
Such a behaviour is expected in a theory with only the triple-Pomeron coupling, and which neglects the more complicated multi-Pomeron vertices, such as the $2 \to 2$ Pomeron coupling and so on \cite{GB}. 

Recall that the two-Pomeron loop gives a positive probability for LRG production but its contribution to the amplitude is negative, see Fig.~\ref{fig:4}. If we now sum up
the diagrams with many two-Pomeron loop insertions (Fig.~\ref{fig:3}(a)) then we `renormalise' the  Pomeron propagator. In this way we obtain the `dressed' propagator in terms of a Schwinger-Dyson equation. The intercept of the final `dressed' propagator
is $\alpha_P(0)=1-\epsilon=1+\Delta-\mbox{`loop'}$. With increasing energy, the available rapidity interval increases, with the result that the loop renormalisation grows and hence the effective intercept $1-\epsilon$ decreases. Asymptotically, the Schwinger-Dyson equation gives
\begin{equation}
\epsilon=\frac{r^2}{2\epsilon}-\Delta,
\label{eq:SD}
\end{equation}
where $r$ is the triple-Pomeron coupling. The loop contribution,  $\mbox{`loop'}=r^2/2\epsilon$, is calculated using the dressed Pomeron propagator, and the factor $1/\epsilon$ comes from the integral over the loop size $\delta y$.\footnote{Strictly speaking, there should also be a Schwinger-Dyson equation giving the dressed triple-Pomeron coupling. However, in this oversimplified estimate, we have truncated the system of 
Schwinger-Dyson equations at the first step.}

For the reasonable values  $r\simeq 0.2$ and $\Delta \simeq 0.3$, the naive estimate based on (\ref{eq:SD}) gives $\epsilon \simeq 0.06$. That is $\alpha^{\rm asympt}_P(0) \simeq 0.94$.
This means that up to a rather large energy ($Y=\ln s<1/\epsilon\sim 18$) the total cross section will grow, but after this, that is just in the LHC region, it will start to decrease. For multiparticle production, such a regime will reveal itself as long-range rapidity fluctuations of the multiplicity arising from two-Pomeron loops which may produce either LRG or a double multiplicity of secondaries in the rapidity domain occupied by the loop.

\section{Saturation of the $k_t$ distribution}
Let us return to the simplified eikonal model (\ref{eq:eik}), written for the partons in the incoming protons. We may ask the question - ``what happens to an incoming fast quark?'' On the one hand, the
factor $\exp(-\Omega)$ describes its absorption. After the inelastic collision the incoming quark disappears from the initial wave function.
On the other hand, the quark cannot just {\it disappear}\footnote{For the gluon the situation is more complicated. Some of gluons can disappear via fusion, $gg\to g$.}. It has baryon charge, etc. Disappearance from the incoming beam (wave function) actually means
{\it migration} -- after the collision the momentum of quark is changed. In the leading logarithmic approximation, the quark mainly changes its transverse momentum. Each new Pomeron-quark coupling may be considered as
a new elastic scattering and it is known that after $n$  scatterings the mean transverse momentum squared, $\langle k^2_t \rangle$, increases $n$ times.

Thus, when we say that unitarity stops the growth of the parton density via an increase of the absorptive correction, actually we mean that the extra partons produced by the cascade are pushed out of the previous domain in $b,\ k_t$ configuration space into a region of larger $k_t$ and $b$. Accounting for the enhanced diagrams we get the same phenomena, not only for the incoming fast partons
(as in the eikonal model), but for any intermediate partons as well.
Thus finally we expect that the particle density will reach saturation at low $k_t$, but will continue to grow with energy at larger $k_t$. Therefore, at higher energies a larger number of minijets with larger $\langle k_t \rangle$ will be produced.  

Due to saturation in the low $k_t$ region at very high energies, the majority of partons will have rather large $k_t$.  We already know that the diffusion in $\ln k_t$ takes place in the perturbative QCD domain \cite{BFKLdiff}, and to provide a smooth matching to the larger $k_t$ domain we need to include some elements of a similar diffusion 
in the description of low (and intermediate) transverse momenta.

\section{Information from double-Pomeron-exchange reactions}

At the Tevatron and the LHC, there is a possibility to study just Pomeron-Pomeron collisions by
selecting events with large rapidity gaps on either side
of some centrally produced hadronic system of mass $M$. This can be done, either, by detecting the two forward outgoing 
protons which each carry a large fraction, $x_L$, of their incoming momenta\footnote{The mass of the centrally produced hadronic system is given by $M^2=s(1-x_{L1})(1-x_{L2})$ and the size of the gaps in rapidity are $\delta y_i=-\ln(1-x_{Li})$.}, say
$x_L>0.96$ , or, just by using the ``gap-hadron-gap'' trigger.
 Since the interaction across each gap is described by 
Pomeron exchange such events may be treated as Pomeron-Pomeron collisions. Usually it is called a Double-Pomeron-Exchange (DPE)
 process.

In a DPE process we may hope to observe the properties of the Pomeron caused by the fact that (i) the Pomeron consists mainly of gluons and (ii) the transverse size of the Pomeron is small, that is the typical transverse momenta, $k_t$, in the diagrams which describe  Pomeron exchange are relatively large. 
Indeed, the value of $k_t$ increases due to BFKL diffusion and screening effects (as explained in the previous Section). 
Another indication in favour of large $k_t$ is the rather small value of the slope $\alpha'\propto 1/\langle k^2_t\rangle$ of the Pomeron trajectory, see Section \ref{sec:trans} and the discussion after eq.(\ref{eq:AFS}). 

For these reasons, DPE events have several characteristic features. First, since the Pomeron consists of mainly of gluons,
we expect there to be a larger fraction  of the glueballs and/or $\eta',\eta$  mesons in the central system $M^2$, than that in $pp$-interactions. Recall that SU(3) iso-singlet $\eta',\eta$ mesons contain a sizeable gluon component. Moreover, because DPE has no ``incoming'' valence quarks, we expect the baryon/hyperon content to be smaller than in $pp$ interactions.

Further information can be obtained by comparing DPE processes with $pp$ interactions at the lower energy $\sqrt{s_{pp}}=M$. For example, the  multiplicity of secondaries is expected to be close to that in the $pp$ case\footnote{We cannot compare with $e^+ e^-$ interactions, since there the multiplicity, and other characteristics of the final distribution, are driven by specific double loagarithms which are different from those in $pp$ collisions \cite{ochs}.}. Strictly speaking, we expect a bit higher multiplicity since in Pomeron-Pomeron collisions the initial energy goes mainly to mesons, while the fraction of baryons is smaller than in $pp$ interactions.

There are several consequences of the small size of the Pomeron. First, the transverse momenta of secondaries are expected to be a bit larger for DPE than that in the $pp$ case at $s_{pp}=M^2$, but smaller (due to the kinematic constraints) than in the original pure inelastic process which takes up the whole initial $pp$ energy $\sqrt{s}$. The high $E_T$ dijets produced by DPE will be dominantly gluon jets.
Next, the relatively small size of the Pomeron should result in a smaller interaction radius, as measured via Bose-Einstein correlations, than that found in the usual inelastic $pp$ collision.
Finally, the probability of double parton scattering, in particular the probability to observe two pairs of high $E_T$ jets, will be larger than that in a $pp$ collision at $s_{pp}=M^2$.

Accounting for the anticipated LHC luminosity, the rate of DPE events is rather large. For example, from Ref. \cite{KMR}, we see that the DPE cross section integrated over the $0.002 <\xi <0.02$ interval,
corresponding to the central production of a system of mass $M\sim$ 100 GeV, is about 10 $\mu b$.  Alternatively, the expected DPE cross section is 
$\xi_1\xi_2d\sigma/d\xi_1d\xi_2\sim 1\ -\ 5\ \mu b$~\cite{KMR}; here $\xi_i=1-x_{Li}$. So it should be possible to study the properties of DPE events in detail.

\section{Conclusions}

We should discuss how the saturation of particle densities can be consistent with confinement. The growth of the transverse momenta $k_t$ tames the increase of the interaction radius with energy. One possibility is that the simultaneous action of confinement and the growth of $k_t$ will finally lead to the saturation of the interaction radius; that is to the saturation of the elastic slope, $B_{\rm el} \to const$ as $s \to \infty$.

\begin{figure} [t]
\begin{center}
\includegraphics[height=8cm]{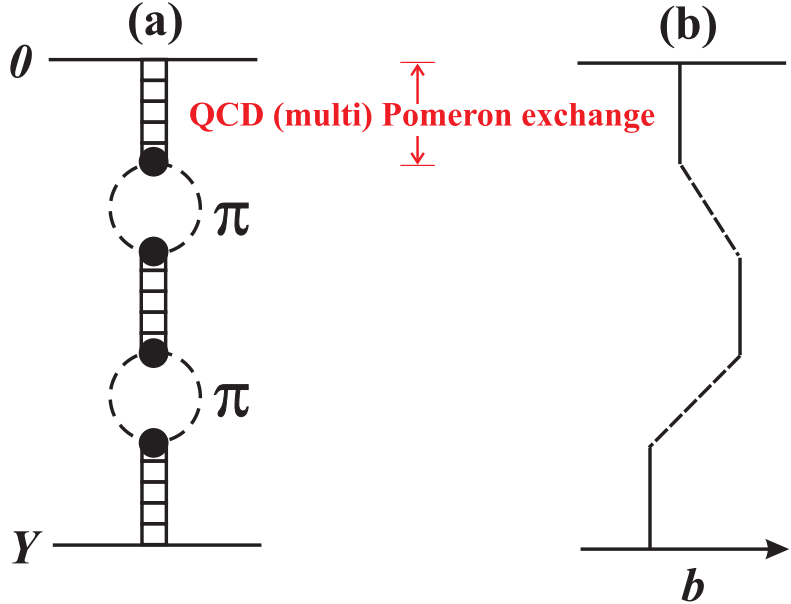}
\caption{The effect of the presence of colourless clusters (meson exchange) on the impact parameter of the produced particles as a function of rapidity.}
\label{fig:M}
\end{center}
\end{figure}

However the idea, discussed at the end of Section \ref{sec:tot}, that when the interaction radius becomes large confinement leads to the formation of colourless clusters, allows an alternative possibility. In a large, but finite, interval of rapidity the evolution is driven by the gluon ladder. In this case the impact parameter is almost frozen; that is, the variation of the interaction radius $\delta b$ is small.  The system then emits a pion (or another meson) and this `colourless cluster' provides a noticeable increase of the impact parameter $b$. Following this, we again have gluons in some interval of evolution, and so on, as shown in Fig. \ref{fig:M}(a). The hard QCD interactions will be responsible for the growth of $k_t$, while the part of the evolution described in terms of colourless mesons will be responsible for the increase in the interaction radius, that is of $B_{\rm el}$, due to the diffusion in {\bf b} plane see Fig. \ref{fig:M}(b). In such a `mixed' picture we expect lower $k_t$ in peripheral collisions (since the available rapidity interval was used to enlarge the value of $b$) and maximal $k_t$ in central collisions. 

It may be informative to study multiparticle production at the LHC for individual events, as well as in terms of inclusive cross sections. At the high LHC energy the multiplicity of secondaries is large. Thus it might be possible to observe the fluctuation of particle densities in individual events like those shown in Fig. \ref{fig:B}, or the effects arising from a mixture of intervals of evolution via colourless meson and coloured gluon exchanges, like those shown in Fig. \ref{fig:M}. In the latter case, the rapidity correlation length (measured via $R_2$ of (\ref{eq:R2})) should be rather small for peripheral collisions since the main part of the available rapidity interval was used for meson exchange (to enlarge $b$) and so only small intervals are left for gluon exchange where 
%we have 
we expect a large probability to have several `cut' Pomerons.

It is appropriate to say a few final words about models of high-energy soft interactions. We emphasize that the aim of experiment is not to reject or to confirm one or another model by saying that the data prefer Monte Carlo version X or Y. Rather, the objective should be to isolate, and to study, the main qualitative features of the interaction. What, therefore, are the requirements of a realistic model of high-energy soft interactions? To describe all the qualitative features discussed in this paper, it is clear that a realistic model should
\begin{itemize}
\item contain, not only the eikonal but also the enhanced multi-Pomeron contributions; 
\item include, not just the triple-Pomeron vertex, but more complicated
multi-Pomeron vertices (otherwise we will get an asymptotically decreasing
cross section); 
\item allow for  `diffusion' of the
partons both in impact parameter, $b$, and in transverse momenta, or, more precisely, 
$\ln k_t$ space.
\end{itemize}
A recent attempt at building such a multi-Pomeron model, tuned to describe all the available data for high-energy soft interactions, is presented in \cite{KMR}. For convenience, we call this the KMR model.

\begin{figure} [t]
\begin{center}
\includegraphics[height=18cm]{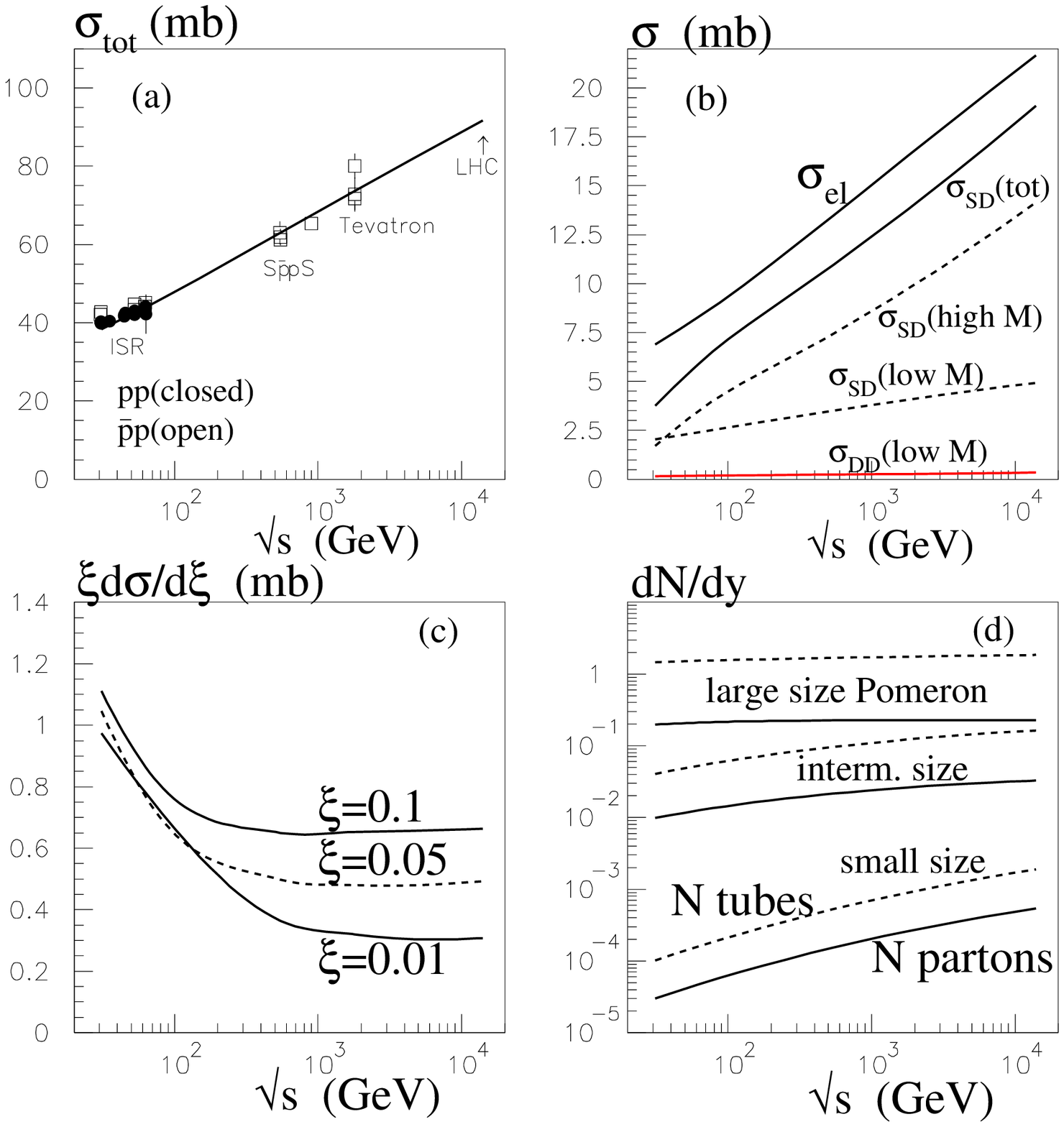}
\caption [*]{The energy dependence of the total (a), elastic and diffractive dissociation (b) $pp$ cross sections, and the cross sections of dissociation to a fixed $M^2=\xi s$ state (c); (d) the parton multiplicity (solid lines) and the number of `colour tubes' (dashed) produced by Pomeron components of different size. The figure is taken from the KMR model of \cite{KMR}.}
\label{fig:KMR}
\end{center}
\end{figure}
Finally, can the characteristic features of the multi-Pomeron description of soft interactions be observed at the high Tevatron and LHC energies? We list some below.  
\begin{itemize}
\item The multi-Pomeron absorptive effects tame the power growth of the $pp$ total cross section leading to values smaller than predicted before. For example, the KMR model prefers a value close to the lower of the two Tevatron measurements, and a value of about 90 mb at the LHC energy of $\sqrt{s}=$ 14 TeV, see Fig. \ref{fig:KMR}(a). At 100 TeV the prediction of 108 mb is at the lower limit of cosmic ray expectations.
\item Simultaneously, these absorptive effects shrink the elastic differential cross section peak as the energy increases through the LHC range more than the expectations arising from a naive effective Pomeron-pole of a Donnachie-Landshoff type of parametrization \cite{DL}.
\item The cross section for proton dissociation into high-mass systems should grow with energy and be comparable to the elastic scattering cross section.  The reason is that although, for each fixed rapidity interval, the probability of high-mass dissociation is relatively small, the overall effect is enhanced by the large phase space that is available in rapidity at high collider energies. The predicted values  of $\sigma_{\rm el}$ and $\sigma_{\rm SD}$ for the KMR multi-Pomeron model are shown in Figs.~\ref{fig:KMR}(b,c) for energies in the Tevatron -- LHC energy range.
\item The main growth in multiplicity, as we go from Tevatron to LHC energies, is due to the small size (`QCD') Pomeron component, which produces particles with typically $p_t \sim 5$ GeV. There is essentially no growth in multiplicity at small $p_t$. This simply confirms the trend that has been observed through the CERN-ISR to Tevatron energy range, see the data points in Fig.~\ref{fig:inclmult}.
\begin{figure} [h] 
\begin{center}
\includegraphics[height=15cm]{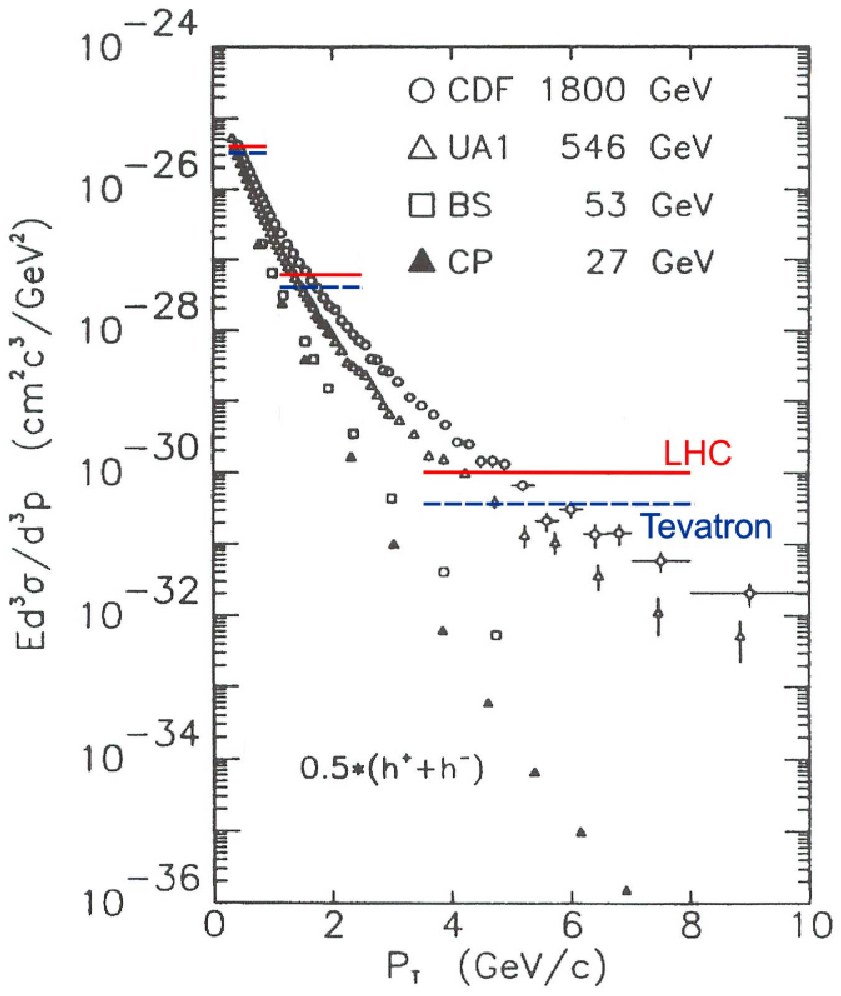}
%\epsf{figure=sumlog.eps,width=14.0cm,height=14.0cm}
%\centerline{\epsfxsize=2.9in\epsfbox{sumlog.eps}}
\caption[*]{The plot is from Ref.~\cite{CDF61}. The horizontal lines, which are superimposed, are the predictions of \cite{KMR} at the Tevatron and LHC energies; the three $p_t$ ranges correspond to the large-, intermediate- and small-size components of the Pomeron, see also Fig.~\ref{fig:KMR}(d)} 
\label{fig:inclmult}
\end{center}
\end{figure}
In other words, starting with the same Pomeron intercept ($\Delta \simeq 0.3$ of Section \ref{sec:PQCD}) the contribution of the large-size
component {\it after the absorptive correction} becomes practically
flat in energy ($\sim s^{0.08}$), while  the small-size contribution, which is much less affected by
the absorption, continues to grow with energy ($\sim s^{0.3}$). Such  behaviour is
consistent with the  experiment (see Fig.~\ref{fig:inclmult}) where the density of low $p_t$
secondaries is practically saturated while probability to produce a hadron
with a large (say, more than 5 GeV) transverse momentum grows with the
initial energy.
\item Multi-Pomeron exchange diagrams have a characteristic pattern of `enhanced' multiplicities of secondary particles, as well as large rapidity gaps, governed by the AGK cutting rules. These lead to long-range rapidity correlations, for example, which may be observed at the LHC via $R_2$ of (\ref{eq:R2}).
\item The observed rate of Large Rapidity Gap (LRG) events in various processes at the LHC will be particularly informative. The survival probability of a LRG may be calculated in terms of a multi-Pomeron exchange model of soft interactions. It is necessary to calculate the survival to both {\it eikonal} soft rescattering (between the colliding protons), and {\it enhanced} rescattering (involving soft interactions with intermediate partons with different $k_t$). To calculate the latter we need to include the $k_t$ dependence of Pomeron exchange, see \cite{KMR} -- this was the third requirement listed for a realistic model.
\item A topical example of a LRG process is the {\it exclusive} production of a heavy mass system $A$, that is $pp \to p+A+p$ where the $+$ signs denote LRGs. This process, with $A=$Higgs, is a novel and promising way to study the Higgs sector at the LHC, which gives a strong motivation for the addition of forward proton detectors to enhance the discovery and physics potential of the ATLAS and CMS detectors at the LHC \cite{epip,albrow}. Already, at the Tevatron, the CDF collaboration have observed exclusive processes with $A=\gamma\gamma$, dijet and $\chi_c$ with rates consistent with the estimates of the LRG survival probabilities predicted by the multi-Pomeron model of soft interactions, see \cite{epip} and references therein. More data on exclusive processes, from both the Tevatron and, especially, the LHC, will be illuminating.
\end{itemize}

\section*{Acknowledgements}

It is a pleasure to thank M.G. Albrow, V.V. Anisovich, A. De Roeck and R. Orava for interesting discussions. MGR would like to thank the IPPP at the University of Durham for hospitality. This work was supported by the grant RFBR
07-02-00023, by the Federal Program of the Russian State RSGSS-3628.2008.2 and by the Russia-Israel grant 06-02-72041-MNTI.

%\newpage

\end{document}